\documentclass[12pt]{article}
\usepackage{amsmath}
\usepackage{graphicx}
\usepackage{times}
\usepackage[T1]{fontenc}
\usepackage[utf8]{inputenc}
\usepackage[margin=1cm]{caption}

\usepackage[square,numbers,sort]{natbib}

\author
{Pankhuri Gupta$^{1\dag}$, Artem Litvinenko$^{2\dag}$, Akash Kumar$^{2,3,4\dag}$,
\\Pranaba Kishor Muduli$^{1\ast}$ and Johan \AA kerman$^{2,3,4\ast}$
\\
\normalsize{$^{2}$Department of Physics, Indian Institute of Technology Delhi, Hauz Khas,}\\
\normalsize{110016, New Delhi, India}\\
\normalsize{$^{2}$Department of Physics, University of Gothenburg, Fysikgränd 3,}\\
\normalsize{412 96, Gothenburg, Sweden}\\
\normalsize{$^{3}$Center for Science and Innovation in Spintronics, Tohoku University, 2-1-1 Katahira,}\\
\normalsize{Aoba-ku, Sendai, 980-8577, Japan}\\
\normalsize{$^{4}$Research Institute of Electrical Communication, Tohoku University, 2-1-1 Katahira,}\\
\normalsize{Aoba-ku, Sendai 980-8577 Japan}\\
\\
\normalsize{$^\dag$These authors contributed equally to this work.} \\
\normalsize{$^\ast$To whom correspondence should be addressed; E-mails:} \\
\normalsize{muduli@physics.iitd.ac.in, johan.akerman@physics.gu.se}
}

\date{}

\begin{document} 

\parindent 0cm
\parskip 12pt

\title{\LARGE\bfseries{Ultra-fast spin Hall nano-oscillator based microwave spectral analysis}} 

\maketitle

\abstract{Ultra-fast spectrum analysis concept based on rapidly tuned spintronic nano-oscillators has been under development for the last few years and has already demonstrated promising results. Here, we demonstrate an ultra-fast microwave spectrum analyzer based on a chain of five mutually synchronized nano-constriction spin Hall nano-oscillators (SHNOs). As mutual synchronization affords the chain a much improved signal quality, with linewidths well below 1 MHz at close to a 10 GHz operating frequency, we observe an order of magnitude better frequency resolution bandwidth compared to previously reported spectral analysis based on single magnetic tunnel junction based spin torque nano-oscillators. The high-frequency operation and ability to synchronize long SHNO chains and large arrays make SHNOs ideal candidates for ultra-fast microwave spectral analysis.}

\section*{Introduction}\label{sec1}
The continuous evolution of communication and signal processing systems~\cite{percival1993spectral,naidu1995modern,Hunter2011ModernSA,Lv2013RTSA} demands the development of ultra-fast spectrum analyzers (SAs) capable of high-frequency analysis combining the highest frequency resolution with sub-$\mu s$ analysis time. Conventional SAs, being swept-tuned or based on Fourier methods, have limitations in terms of speed of analysis, complexity, and computational requirements. In recent years, significant progress has been made in utilizing spin-torque nano-oscillators (STNOs) for ultra-fast spectrum analysis~\cite{louis2018ultra, litvinenko2020ultrafast, litvinenko2022ultrafast}. The use of such nanoscale devices offers advantages such as low power consumption~\cite{LeeSTNOpowerEfficiency2019,litvinenko2021analog,Jiang2024Spin}, compatibility with CMOS technology~\cite{ma2021microwave}, wide frequency tuning range~\cite{litvinenko2022ultrafast}, and very rapid frequency tunability~\cite{QuinsatSTNOmodulation2014,Ruiz-CalaforraSTNOs2017,sharma2017high} due to the small size and hence inherently low parasitic capacitance and inductance affecting the frequency response time. Earlier experimental studies demonstrated the potential of STNOs in achieving fast and broadband spectrum analysis with a resolution bandwidth primarily defined by the bandwidth theorem when sweep frequencies are above the STNO generation linewidth~\cite{litvinenko2020ultrafast, litvinenko2022ultrafast}. Magnetic tunnel junction (MTJ) based vortex-state STNOs exhibit an amplitude relaxation frequency of 2-3 MHz and a linewidth of hundreds of kHz, allowing sweeping frequencies from 50 kHz up to 1.5 MHz with a corresponding frequency resolution bandwidth of spectrum analysis~\cite{litvinenko2020ultrafast}. However, the SA frequency range was limited to about 300 MHz due to the low auto-oscillation frequency of vortex-state STNOs. This limitation was later solved by exploiting a uniform-state STNO operating in a frequency range between 8.6 and 9.6 GHz~\cite{litvinenko2022ultrafast}. However, while the frequency range was much improved, the frequency resolution deteriorated to 35 MHz due to the poor frequency stability of uniform-state STNOs.

In an attempt to combine a high and wide frequency range with best-in-class frequency resolution, we here explore mutually synchronized nano-constriction spin Hall nano-oscillators~\cite{Demidov2012,demidov2014nanoconstriction,duan2014nanowire,behera2022energy,rajabali2023injection,behera2024ultra} (NC-SHNOs) for ultrafast microwave signal analysis. In NC-SHNOs, a direct charge current flows through a NC in a heavy metal/ferromagnetic (HM/FM) bilayer. The spin Hall effect~\cite{Dyakonov1971,Hirsch1999,Sinova2015Rev} in the HM layer generates a transverse spin current into the FM layer, which exerts an anti-damping torque on its magnetization~\cite{Liu2012PRL}. Above a certain threshold current, the FM magnetization in the NC starts to auto-oscillate, creating a microwave voltage via the anisotropic magnetoresistance (AMR)~\cite{Demidov2012,duan2014nanowire,divinskiy2017nanoconstriction}. 

The performance of NC-SHNOs has seen remarkable progress in the last few years thanks to mutual synchronization in chains~\cite{awad2017long,kumar2023robust,litvinenko2023phase,kumar2024spin} and two-dimensional arrays~\cite{zahedinejad2020two}. This has led to new NC-SHNO applications in emerging fields, such as neuromorphic computing~\cite{zahedinejad2020two,zahedinejad2022memristive,sethi2023compensation} and Ising machines~\cite{houshang2022SHNOIM,gonzalez2024spintronic}. In particular, mutual synchronization of SHNOs improves three key metrics important for ultrafast SAs: \emph{i}) the linewidth decreases linearly with the number $N$ of mutually synchronized SHNOs, leading to higher frequency resolution; \emph{ii}) the power increases with $N$, improving the SNR; and \emph{iii}) the non-linearity factor increases due to reduced radiative energy losses~\cite{kumar2023robust}, increasing the frequency range and tunability. As a consequence, utilizing mutually synchronized SHNO chains with five nano-constrictions, we here demonstrate ultra-fast spectral analysis in the 10 GHz range with a resolution that is an order of magnitude better than previously reported using uniform-state STNOs~\cite{litvinenko2022ultrafast}. 

\begin{figure*}[b!]
    \centering
    \includegraphics[width=8cm]{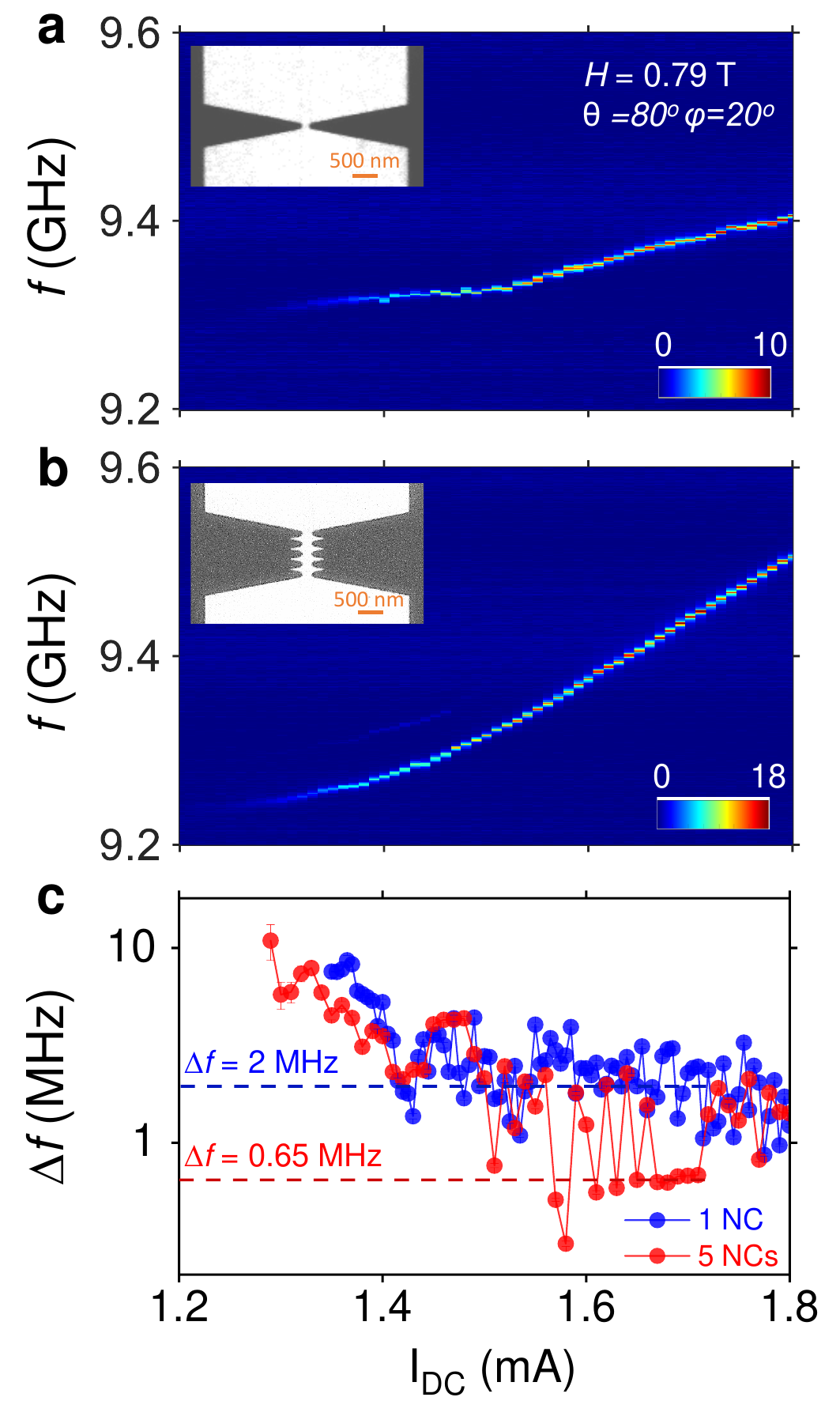}
    \caption{\textbf{Free-running properties of spin Hall nano-oscillator chains:} Power spectral density (PSD) for (\textbf{a}) single nano-constriction (1 NC) SHNO and    
    (\textbf{b}) five nano-constriction (5 NCs) SHNOs in a chain. 
    (\textbf{c}) Spectral linewidth for 1 NC and 5 NCs. The measurements are performed at external magnetic field, H = 0.79 T applied at out-of-lane angle, $\theta= 80^\circ$ and in-plane angle, $\phi = 20^\circ$
    } 
    \label{Fig1} 
\end{figure*}

\section*{The SHNOs and their characterization}\label{sec2}
We utilize five mutually synchronized SHNOs fabricated using W (5 nm)/NiFe (3 nm)/Al$_{2}$O$_{3}$ (4 nm), as shown in Fig.~\ref{Fig1}. Details of the fabrication and mutual synchronization of SHNO chains are discussed in the method section. The microwave voltage signal is generated via the AMR effect. The SHNOs are then first characterized for their free-running properties. Figure~\ref{Fig1}a$\&$b shows the auto-oscillation power spectral density (PSD) for a single and five NC-SHNOs, respectively; the inset shows the scanning electron microscopy (SEM) image of the actual devices. The five NCs show robust mutual synchronization and deliver both much larger output power (18 \emph{vs.}~10 dB/noise) and much narrower linewidth (0.65 \emph{vs.}~2 MHz; Fig.~\ref{Fig1}c) than a single NC. In addition, the mutually synchronized SHNO chain shows a much larger frequency tunability of $df/dI=$ 0.67 \emph{vs.}~0.3 GHz/mA, measured between 1.5 and 1.8 mA. The frequency range can be further extended to $2-30$~GHz via a magnetic field and its angular orientation (as shown in Supplementary Note 1)~\cite{Zahedinejad2018apl}. The higher output power, the lower linewidth, and the greater frequency tunability are all three of utmost importance for ultra-fast spectrum analysis.

\begin{figure*}[ht!]
    \centering
    \includegraphics[width=16cm]{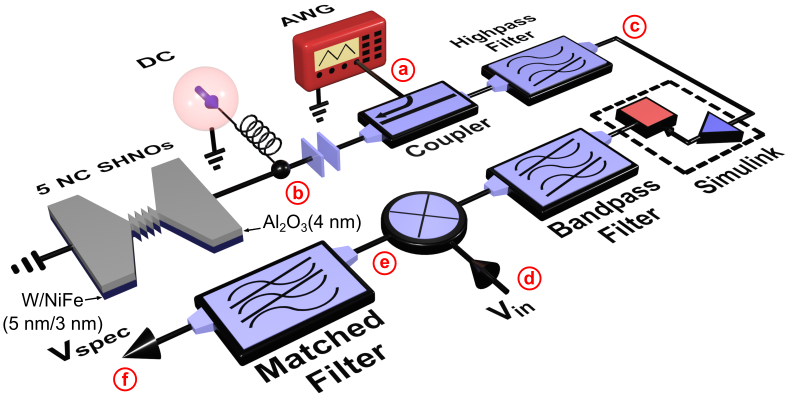}
    \caption{\textbf{NC-SHNO-based fast spectrum analysis.} Schematic of the NC-SHNO-based spectrum analyzer consisting of two blocks: an NC-SHNO-based sweep-tuned reference signal generator and a signal processing block with a mixer and a matched filter. DC current source injects bias current into the SHNO chain at point (b). RF signal goes to a highpass filter after the DC block capacitor and coupler. A coupler injects a triangle-shaped signal at point (a) to modulate the SHNO chain frequency. The filtered signal after a highpass filter gets digitized and fed to the Simulink environment. Additional filtering with a bandpass filter is done to improve the SNR of the SHNO frequency chirp signal. An unknown signal is added at point (d) to be mixed with the SHNO frequency chirp signal. A resulting intermediate signal at point (e) is postprocessed with a matched filter to perform inverse Z-transform and recover the spectral components of the signal $V_{in}$ in the form of voltage spikes.} 
    \label{Fig2}
\end{figure*}

\section*{Ultra-fast spectrum analyzer scheme}\label{sec3}
The scheme of the experimental NC-SHNO-based ultra-fast SA is shown in Fig.~\ref{Fig2}. It consists of signal generation and processing sections. For signal generation, a constant $I_{\rm DC}$ is applied to the NC-SHNO to drive a microwave auto-oscillation signal of a constant carrier frequency $f_{\rm 0}$. An arbitrary waveform generator (AWG) is then connected via a microwave coupler to modulate the SHNO frequency with a voltage $V_{\rm sw}$ having a triangular shape in time and a frequency $f_{\rm sw}$ = 1/T. $V_{\rm sw}$ is adjusted so that the auto-oscillation signal generated by the NC-SHNO ($V_{\rm SHNO}$) lies within the linear frequency range. The SHNO signal is passed through a high pass filter to remove any spurious harmonics of the triangular signal and then amplified into the signal $V_{\rm ref}$ having frequency chirp $f_{\rm ref}$(t). 
$V_{\rm ref}$ is digitized by a real-time oscilloscope and processed within a MATLAB environment using Simulink. 

In the second section, signal processing and spectrum analysis of an external signal $V_{\rm in}$ is performed. The reference signal $f_{\rm ref}$(t) is first filtered with a narrow-band bandpass filter, [9.38 -- 9.44] GHz, whose bandwidth corresponds to the frequency modulation range of the NC-SHNO $\Delta F_{\rm sw}$ to improve the signal-to-noise ratio. After filtering, it is mixed with the external signal $V_{\rm in}$ of unknown frequency to be analyzed. The resulting intermediate frequency signal $V_{\rm if}$ is passed through a matched filter with a predefined amplitude-frequency response. A matched filter performs a convolution operation on an intermediate frequency signal, $V_{\rm if}$, which results in an output spiking signal, $V_{\rm spec}$, where the temporal position of the voltage peak is proportional to the frequency of the input signal. 

The temporal width of the peak ${\tau}$ in the voltage signal, $V_{\rm spec}$ and the corresponding frequency resolution bandwidth (RBW) of the spectrum are bound by:
\begin{equation}
    \centering
     {\Delta}F_{\rm RBW}= {\Delta}F_{\rm sw} \frac{{\tau}}{T}
    \label{eq:RBW}   
\end{equation}

\begin{figure*}[ht!]
    \centering
    \includegraphics[width=10cm]{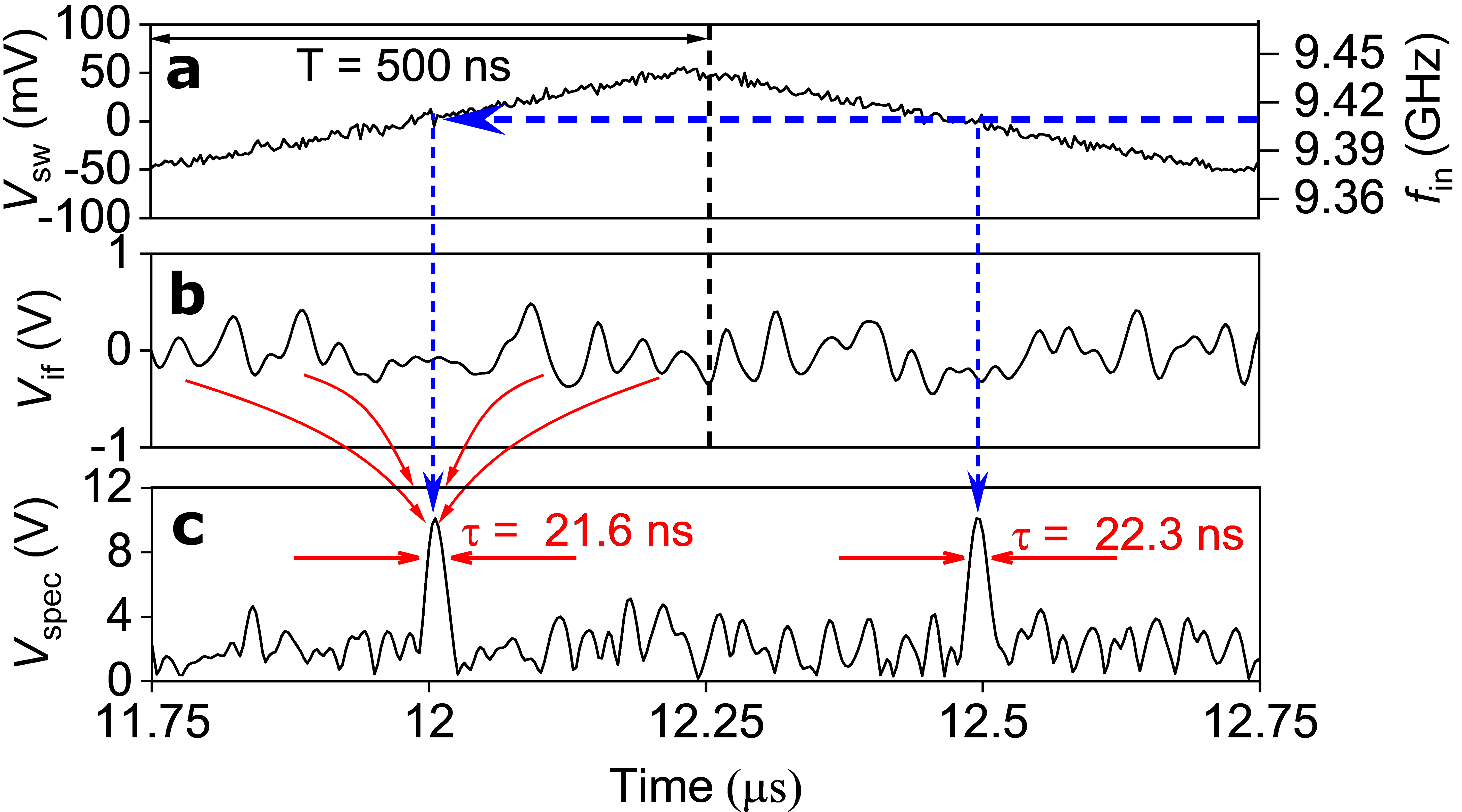}
    \caption{\textbf{Spectral analysis of a single-tone external sinusoidal signal ($f_{\rm in}$= 9.41~GHz).} (\textbf{a}) sweeping signal $V_{\rm sw}$ (measured at point (a) in Fig.2a.) (\textbf{b}) signal $V_{\rm if}$ (measured at point (e) in Fig.2a.) (\textbf{c}) signal $V_{\rm spec}$ (measured at point (f) in Fig.2a.). 
    } 
    \label{Fig3} 
\end{figure*}

\section*{Results}\label{sec4}
\subsection*{Principle of operation}
Fig.~\ref{Fig3} shows the analysis of a pure sinusoidal signal $V_{\rm in}$  ($f_{\rm in}$ = 9.41 GHz, amplitude 1 V): The top panel represents the sweeping triangle signal $V_{\rm sw}$, with frequency $f_{\rm sw}$ = 1/($T$) = 2 MHz, with the right-hand side $y$ axis showing the frequency of the resulting SHNO signal, or $f_{\rm ref}$(t); the middle panel represents the intermediate frequency signal $V_{\rm if}$ resulting from the mixing of the chirped signal $V_{\rm ref}$ and external signal $V_{\rm in}$; the bottom panel represents the output signal from the matched filter $V_{\rm spec}$.
In the output signal, the narrow and high amplitude peak in $V_{\rm spec}$, even with noisy $V_{\rm sw}$ and $V_{\rm if}$, shows the importance of the matched filter. 
The purpose of the matched filter is to perform a convolution procedure over the intermediate frequency signal $V_{\rm if}$, which drastically maximizes the effective signal-to-noise ratio (SNR) by squeezing the frequency swept signal contained in $V_{\rm if}$ into a narrow voltage peak whose temporal width is limited by Eq.~\ref{eq:RBW}, while the uncorrelated noise passes through the filter unchanged. The working principle of the matched filter is explained in Supplementary Note 2. The SNR improvement depends on the squeezing factor and is defined by the ratio between the theoretically minimal temporal width of the peak $\tau$ and the sweeping time. The SNR improvement of the matched filter can also be defined as the ratio between the frequency modulation range ${\Delta}F_{\rm sw}$ of the NC-SHNO  and the sweeping frequency $f_{\rm sw}$:

\begin{equation}
    \centering
     \mathit{SNR}_{\rm filter}=\frac{T}{{\tau}}=\frac{{\Delta}F_{\rm sw}}{f_{\rm sw}}
    \label{eq:SNR} 
\end{equation}
\\
For ${\Delta}F_{\rm sw}$ = 60 MHz, $f_{\rm sw}$ = 2 MHz and 1 MHz, the SNR improvement from the filter equals 30 and 60, respectively. Hence, despite the low SNR of the intermediate signal $V_{\rm if}$, the output signal $V_{\rm spec}$ accurately represents the input signal spectrum with a high SNR between the peak amplitude and the noise floor.

Finally, the bottom panel of Fig.~\ref{Fig3} has a characteristic duration of about $\tau=$~21.6~ns, which is correlated to a frequency resolution bandwidth of $\Delta F_{\rm RBW}=$~2.59~MHz.

\subsection*{Resolution bandwidth}

\begin{figure*}[ht!]
    \centering
    \includegraphics[width=16cm]{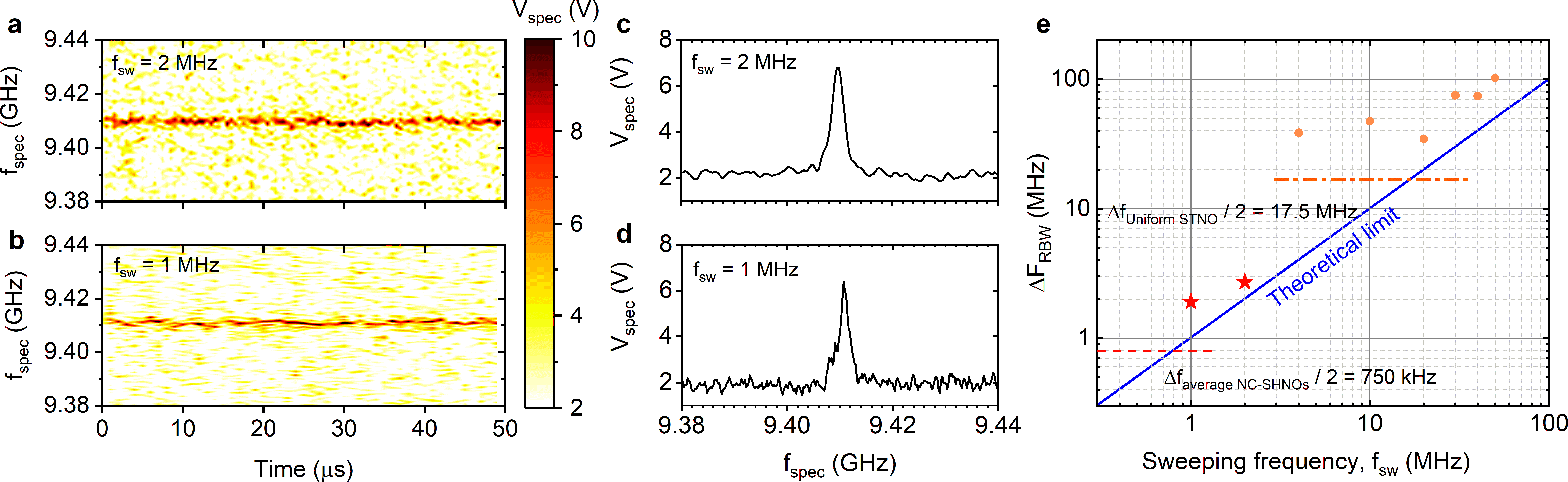}
    \caption{\textbf{Experimental results of the spectrum analysis and resolution bandwidth.}
    The spectrum analysis results ($V_{\rm spec}(t)$) of external signal with different sweeping frequencies $f_{\rm sw}$ = 2~MHz (\textbf{a}) and 1~MHz (\textbf{b}) .
    Output spectrum $V_{\rm spec}$. Statistics are performed on 10 random consequent measurements at $f_{\rm sw}$ = 2 MHz (\textbf{c}) and 1 MHz (\textbf{d}) . 
    (\textbf{e}) Resolution bandwidth (RBW) of the spectrum analysis  ${\Delta}F_{\rm RBW}$ as a function of the sweeping frequency $f_{\rm sw}$. The solid blue line indicates the theoretical limit ${\Delta}F_{\rm RBW}= f_{\rm sw}$. The red dashed line shows the practical limit imposed by an averaged half-linewidth of NC-SHNOs. The red dash-dotted line shows the practical limit imposed by an averaged half-linewidth of the uniform-state STNO.
    } 
\label{Fig4} 
\end{figure*}

In Fig.~\ref{Fig4}a$\&$b, we demonstrate the results of experimentally obtained spectrum analysis $V_{\rm spec}(t)$ of an external signal for two different sweep frequencies $f_{\rm sw} =$ 2~MHz (Top panel) and 1~MHz (Bottom panel). While the frequency resolution at $f_{\rm sw} =$ 1~MHz is better than at 2~MHz, at 
1~MHz, the absolute frequency resolution becomes more affected by the 1/$f$-noise of the NC-SHNO~\cite{litvinenko2023phase}, which results in a drift of the identified central frequency at $\mu s$ time scale. In Fig.~\ref{Fig4}c$\&$d, we show an output signal $V_{\rm spec}$ averaged over 10 random consecutive measurements at $f_{sw}$ = 2~MHz (Top panel) and $f_{sw}$ = 1~MHz (Bottom panel)  for an input frequency of $f_{\rm in}$ = 9.41~GHz. 

Fig.~\ref{Fig4}e shows the absolute frequency resolution bandwidths obtained using NC-SHNO-based SA at different sweep rates $f_{sw}$. The solid blue line indicates the theoretical limit obtained from the “bandwidth” theorem: ${\Delta}F_{\rm RBW}= f_{\rm sw} = 1/T$. Here, the orange circles show the results from the previous study performed using a uniform-state STNO operating in the GHz range~\cite{litvinenko2022ultrafast}. The dashed red line and dash-dotted red lines show the NC-SHNO and uniform-state STNO averaged half-linewidths, respectively. We note that since the NC-SHNO is swept in a wide range of current, where its linewidth changes significantly, an averaged half-linewidth should be taken to estimate the practical limit for the lowest achievable resolution bandwidth ${\Delta}F_{\rm RBW}$. The best frequency resolution bandwidth achieved by NC-SHNO-based SA is 1.9~MHz, which is 20 times better than that obtained with a uniform-state STNO. It is important to emphasize that such tremendous improvement in the frequency resolution using NC-SHNO comes from its particularly low linewidth of ${\Delta}f_{\rm NC-SHNO}$ = 800~kHz.

\subsection*{Ultra-fast spectrum analysis of time-varying signals}
\begin{figure*}[ht!]
    \centering
    \includegraphics[width=8cm]{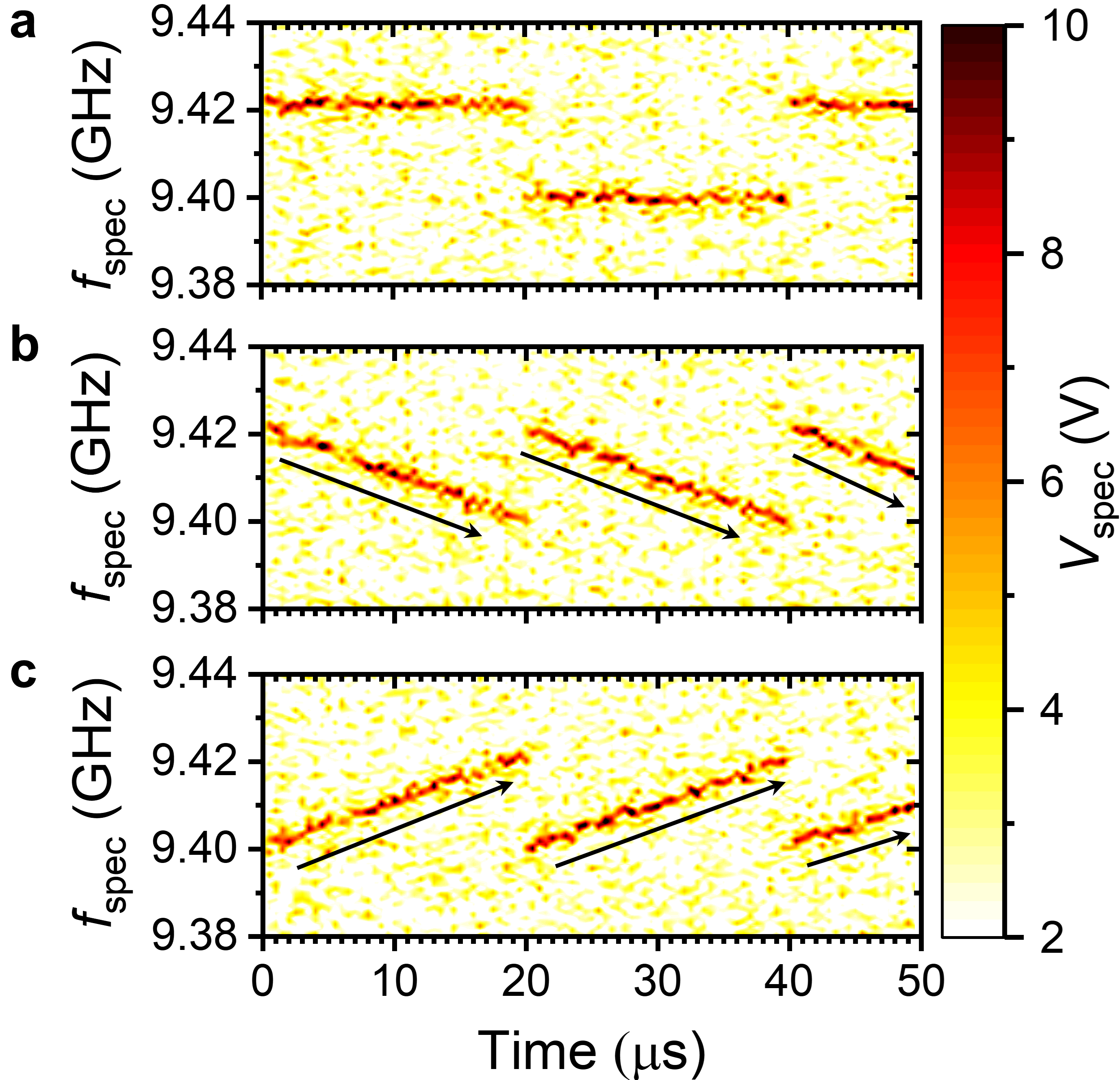}
    \caption{\textbf{Signals with various modulation patterns.}
    (\textbf{a}) Signal modulation with the shift-keyed frequency between 9.40 and 9.42 GHz. (\textbf{b}) Signal modulation with frequency varies discontinuously in a sawtooth fashion with increasing frequency. (\textbf{c}) signal modulation with frequency varies discontinuously in a sawtooth fashion with decreasing frequency. 
    } 
    \label{Fig5} 
\end{figure*}

We finally demonstrate SA of an external signal $V_{\rm in}$ with time-varying frequency components $f_{\rm in}(t)$, which shows the potential of the SHNO-based sweep-tuned spectrum analyzer. The three panels in Fig.~\ref{Fig5} demonstrate the experimental results of the frequency analysis of three external input signals. In the top panel, the frequency experiences discontinuous shift keying, while the middle panel shows sawtooth-like temporal variations with increasing frequency, and the bottom panel shows sawtooth-like temporal variations with decreasing frequency.

\section*{Discussion}\label{sec5}
The exceptionally narrow linewidth of mutually synchronized NC-SHNOs allowed us to achieve ultra-fast spectral analysis with one order of magnitude better frequency resolution bandwidth compared to the previously reported uniform-state STNOs. The NC-SHNO-based spectrum analyzer also operates at an operating frequency of one order of magnitude higher than vortex-based STNOs. In the current implementation, we chose an operating bandwidth of 60~MHz to keep a reasonable SNR of the intermediate signal $V_{\rm if}$. Nevertheless, there is potential for a substantial increase up to GHz bandwidths. In our current demonstration, we used five mutually synchronized NC-SHNOs with a signal power of 18 dB over the noise floor. Recently, robust mutual synchronization in long chains of up to 21~NC-SHNOs was successfully demonstrated~\cite{kumar2023robust} with a signal power density of more than 30~dB over noise, with an integrated power of 300~pW. The operating frequency was around 18 GHz with a tuning bandwidth of over a GHz. Over 2 GHz of voltage-gated tuning of SHNOs operating at 12--14 GHz was also recently demonstrated~\cite{kumar2022fabrication}. These results suggest a dramatic further potential for ultrafast spectral analysis at yet higher frequencies and over wide frequency ranges. Moreover, an increase in the sweeping bandwidth from 60~MHz to 1-2~GHz allows a corresponding increase in the sweep frequency $f_{\rm sw}$ of  30-60~MHz.

We would also like to note that the matched filter can potentially be implemented in the form of an analog spinwave-based dispersive delay line, and the mixing part may be implemented using the SHNO nonlinearity. Thus, future progress could potentially demonstrate an all-spintronic ultra-fast spectrum analyzer. Such a fully spintronic analyzer would be of interest for space applications where radiation hardness is important and the use of CMOS should be limited.

\section*{Conclusion}\label{sec6}
We demonstrate ultra-fast spectral analysis using mutually synchronized nano-constriction spin Hall nano-oscillators. The observed frequency resolution bandwidth of 1.9~MHz is one order of magnitude better than previously reported ultra-fast spectral analysis using uniform-state spin torque nano-oscillators. Compared to vortex-state-based STNO spectrum analyzers, NC-SHNO-based spectrum analysis demonstrates a one-order higher operating frequency. Moreover, NC-SHNO-based spectrum analyzers show prospects for ultra-wideband analysis due to their very wide frequency tunability. In addition, NC-SHNOs can be successfully used for ultra-fast time-resolved spectrum analysis of signals with fast-changing frequency components.

\section*{Methods}\label{sec7}
\subsection*{Spin Hall nano-oscillator fabrication}
The thin film stacks of W (5 nm)/NiFe (3 nm)/Al$_2$O$_3$ (4 nm) are prepared using DC/RF magnetron sputtering (AJA Orion 8) on a high resistive Si substrate ($\rho>$10,000 $\Omega-cm$). The SHNOs devices with 150 nm widths are fabricated using e-beam lithography (Raith EBPG 5200) followed by Ar-ion milling.~\cite{kumar2022fabrication} The chain of five oscillators is fabricated by placing five nano-constrictions separated by 200 nm using a wider bridge between them. The 200 nm separation allows a strong coupling between extended modes of auto-oscillations in the SHNOs, leading to robust mutual synchronization.~\cite{kumar2023robust} The Ground-Signal-Ground (G-S-G) contact pads are fabricated in a subsequent step using mask-less UV lithography and lift-off technique. Deposition of Cu (800 nm)/Pt (20 nm) for contact pads is performed using DC magnetron sputtering.

\subsection*{Electrical measurements}
The electrical measurements for the characterization of free-running properties of SHNO devices are performed using a custom-built G-S-G pico-probe setup (150 $\mu$m pitch from GGB industries) between the electro-magnet poles. The sample stage has the functionality of motorized out-plane rotation. A DC current is supplied to the SHNO devices using a current source (KE 6221). The magnetic field of 0.79 T is applied at $\theta = $80$^\circ$, out-of-plane, and $\varphi = $20$^\circ$, in-plane angles to achieve extended modes in NiFe thin films.~\cite{dvornik2018origin} The generated RF auto-oscillations are observed using spectrum analyzer/oscillators (R$\&$S FSV) after pre-amplification using low-noise amplifiers. The anisotropic magneto-resistance measurements are performed by in-plane scanning of the applied magnetic field (0.1 T) using a vector magnet (GMW).

\section*{Acknowledgements}
This work was supported by the Horizon 2020 research and innovation programme, ERC Advanced Grant No.~835068 "TOPSPIN" and the ERC Proof of Concept Grant No.~101069424 "SPINTOP".

\section*{Author contributions}
A.L. designed the circuit; A.K. designed and nanofabricated the NC-SHNOs; P.G., A.L. and A.K. performed the measurements and analyzed the data; J.Å. managed the project; all co-authors contributed to the manuscript, the discussion, and the analysis of the results.

\bibliography{references}

\end{document}